\documentstyle[twoside,psfig]{acta96}

\begin{document}

\prvastrana=381
\poslednastrana=6

\def\autor{M. Fortunato et al}
\def\nazov{Quantum control of chaos}

\headings{381}{386}
\setcounter{page}{381}

\title{Quantum Control of Chaos inside a Cavity}

\author{M.Fortunato\footnote{\email{for@physik.uni-ulm.de}},
        W. P. Schleich\footnote{Also at Max-Planck Institut
        f\"ur Quantenoptik, D-85748 Garching bei M\"unchen, Germany}}
{Abteilung f\"ur Quantenphysik, Universit\"at Ulm,
          Albert-Einstein-Allee 11 \\ D-89069 Ulm, Germany}

\author{G. Kurizki}
{Department of Chemical Physics, The Weizmann Institute of Science \\
Rehovot 76100, Israel}

\datumy{31 May 1996}{7 June 1996}
\abstract{By sending many two-level atoms through a cavity resonant
          with the atomic transition, and letting the interaction times
          between the atoms and the cavity be randomly distributed, we end
          up with a predetermined Fock state of the electromagnetic field
          inside the cavity if we perform after the interaction with the
          cavity a conditional measurement of the internal state of each atom
          in a coherent superposition of its ground and excited states.
          Differently from previous schemes, this procedure turns out to be
          very stable under fluctuations in the interaction times.}

\kapitola{1. Introduction}

In the last decade, a great deal of attention has been dedicated to
the problem of quantum state preparation [1-5], since the
availability of non-classical states can allow the investigation of
fundamental problems in Quantum Mechanics [1]. Among them, Fock
states [2] are particularly intriguing because they do not
present intensity fluctuations.
Two major approaches to achieve this goal have been
proposed: the first one is based on unitary evolution [3],
that is on finding the right Hamiltonian which
evolves the initial state to the desired final one. The second approach is
based on the conditional measurement (CM) scheme [4] in which
the desired state is achieved after a measurement is performed on one of
two interacting systems. The CM approach has the disadvantage
that unsuccessful runs (experiments
in which the measurement does not give the right result) must be discarded,
and therefore it has a success probability which is always less than unity.
On the other hand, it has the clear advantage of a simple Hamiltonian
evolution, as, for example, the Jaynes-Cummings model.

In this paper, we present a new scheme which---differently from previous
ones [5]---allows the preparation of Fock states inside a
cavity in the presence of even large fluctuations in the interaction times
between the two-level atoms and the cavity field. It is based on the
CM approach and on the quantum interference between the two possible
final states of the atom. The presence of both these effects performs
a strong suppression of the fluctuations in the atomic velocities, and
makes the convergence of the photon-number distribution towards that of
a number state possible.

Our proposal connects as well to the recently introduced field of
chaos control [6]. In fact, it has been shown theoretically
and experimentally that it is possible to use the extreme sensitivity
of chaotic systems to stabilise regular periodic orbits in the chaotic
dynamics. The classical version of our model, implemented
via non-selective measurements (NSMs), is indeed chaotic even for a
small spread in the interaction times [5].
Our CM scheme could then be interpreted as a ``quantum way'' of
controlling chaos. In this view, our method is a new scheme
which can effectively restore fixed points in the quantum dynamics
of a classically chaotic system.

\kapitola{2. The model}

We consider a model in which many two-level atoms are sent through
a cavity whose frequency is resonant with the atomic transition.
The atoms cross the cavity sequentially (one at a time) so that at most
one atom is present inside the cavity.
In the general case, the atoms are initially prepared in a coherent
superposition of their ground and excited states [7] with
the help of two classical fields $E_1$ (resonant) and $E_2$ (non-resonant).
After the preparation, the state of the $k$th atom is
$
|\phi_k^{(i)}\rangle = \alpha^{(i)}_k |e\rangle + \beta^{(i)}_k|g\rangle\; ,
$
where $|g\rangle$ and
$|e\rangle$ are the ground and the excited state of the atom, respectively.
On the other hand, the cavity field is initially prepared by a classical
oscillator $E_3$ in a coherent
state
\begin{equation}
|\psi_0\rangle = \sum_{n=0}^{\infty} d_n^{(0)} |n\rangle =
\exp\left(-{|\alpha|^2 \over 2}\right) \sum_{n=0}^{\infty}
{\alpha^n\over \sqrt{n!}} |n\rangle\; .
\label{eq:infstate}
\end{equation}

The interaction between the atoms and the cavity is described [8]
by the resonant Jaynes-Cummings model, namely, the total Hamiltonian 
of the system (atom and field) is given by
\begin{equation}
\hat{H} = \frac{1}{2}\hbar\omega\hat{\sigma}_z+\hbar\omega
\left(\hat{a}^\dagger\hat{a}\right)+\hbar g\left(\hat{a}^\dagger
\hat{\sigma}_-+\hat{a}\hat{\sigma}_+\right)
 = \hat{H}_0+\hat{H}_{\rm int}\; ,
\label{eq:totham}
\end{equation}
where $\omega$ is the resonance frequency of the atoms and of the cavity,
$\hat{a}$ and $\hat{a}^\dagger$ are the usual annihilation and creation
operators for the field mode, $\hat{\sigma}_i$ are the Pauli operators and
$
\hat{H}_{\rm int} = \hbar g\left(\hat{a}^\dagger
\hat{\sigma}_-+\hat{a}\hat{\sigma}_+\right)\;.
$
In Eq.~(\ref{eq:totham}) $g$ denotes the coupling constant between the
atoms and the field mode.
The atoms are detected, after they have passed through the cavity,
in the coherent superposition
$
|\phi^{(f)}_{k}\rangle = \alpha^{(f)}_k |e\rangle +
                            \beta^{(f)}_k|g\rangle\;,
$
again thanks to two classical fields with which
the atoms interact after they exit the cavity: $E_4$ (non-resonant)
and $E_5$ (resonant), like in the preparation region but in
reverse order [7].

The problem is such that it can be treated iteratively, finding the
recurrence relation between the coefficients of the Fock basis
expansion of the field state inside the cavity after the
interaction and the conditional measurement of the $k$th
atom and the corresponding coefficients before
[after the $(k-1)$th atom]. Then, by repeatedly applying such a recurrence
relation, we can compute the coefficients in the number basis of the
final field state (after a sequence of $N$ atoms), starting from the
initial coherent state, Eq.~(\ref{eq:infstate}). 
For convenience we will work in the
interaction picture [where $H_{\rm int}$ is regarded as the interaction
part of the Hamiltonian~(\ref{eq:totham})], and we will assume that
the resonant fields $E_1$, $E_3$, and $E_5$ are phase-locked.
In what follows we neglect spontaneous emission (since the transit
time of the atoms is much smaller than the typical decay time) and
any dissipation inside the cavity, assuming that the time required
for the whole sequence of atoms is much smaller than the cavity
lifetime.

Computing the evolved atom-field entangled state through the
unitary evolution given by the Hamiltonian~(\ref{eq:totham}),
and then projecting it onto the final atomic state, the following
recurrence relation between the state of the field
$|\psi_k\rangle=\sum_n d_n^{(k)}|n\rangle$ after the $k$th atom
and the corresponding state after the $(k-1)$th atom can be found
\begin{eqnarray}
d_n^{(k)} = P_k^{-1/2}\left\{ \left[
\alpha_k^{(i)}\alpha_k^{(f)*} C_n^{(k)}\right.\right.
 & + & \left.\left.\beta_k^{(i)}\beta_k^{(f)*}C_{n-1}^{(k)}\right]
       d_n^{(k-1)}
 - i\alpha_k^{(i)}\beta_k^{(f)*}S_{n-1}^{(k)} d_{n-1}^{(k-1)}
 \right.
\nonumber \\
 & - & \left. i\beta_k^{(i)}\alpha_k^{(f)*}S_n^{(k)} d_{n+1}^{(k-1)}\right\}\;,
\label{eq:transf}
\end{eqnarray}
where $C_n^{(k)}=\cos(g\tau_k\sqrt{n+1})$,
$S_n^{(k)}=\sin(g\tau_k\sqrt{n+1})$, and $P_k$ is the
success probability of the CM, which is given by
the norm of the projection onto the final atomic state.
In Eq.~(\ref{eq:transf}) it is understood that $d_{-1}^{(k-1)}=0$.

\kapitola{3. Field state dynamics in the presence of random fluctuations}

In this section we study the behaviour of the final field state
(after many atoms have passed through the cavity) when we
allow a spread in the atomic velocities, that is in the interaction
times of the atoms with the cavity.
The JC model has already been proposed [5] for the
production of Fock states of the electromagnetic field inside a
cavity, in connection with NSMs.
That model, however, is very sensitive to even a small spread in
atomic velocities [5], which eventually makes the system
escape any fixed points in the evolution of the photon number distribution.
As a consequence, such a scheme---notwithstanding its great pioneering
value---is of no practical use in the production of Fock states, since
any velocity selector for atomic beams allows a spread in the atomic
velocities. In that approach, the convergence to a Fock state is due to
the existence of the well known ``trapping states'' in
the Jaynes-Cummings evolution [8]. However, in the case of
$|e\rangle \rightarrow |e\rangle$ (elastic CMs) or $|e\rangle\rightarrow
|g\rangle$ (inelastic CMs) schemes, if the interaction times $\tau_k$
fluctuate randomly with $k$, there is a critical value of the spread
$\Delta\tau$ above which the number
distribution will broaden rather than converge. We can estimate this
critical value $\Delta\tau_c$ as the difference between the trapping
and the anti-trapping interaction times for a given $n$ [8], namely,
\begin{equation}
\Delta\tau_c\cong{\pi\over g \sqrt{n+1}}-{\pi\over 2 g \sqrt{n+1}}
={\pi\over 2 g \sqrt{n+1}}\; ,
\label{eq:dtcrit}
\end{equation}
where the sub-ensemble of $|e\rangle \rightarrow |e\rangle$
CMs has been considered.
This phenomenon is shown in Fig.~1, where we plot for
$\Delta\tau=\Delta\tau_c$ the mean
value $\langle n \rangle$ and the rms spread $\Delta n =(\langle n^2
\rangle - \langle n \rangle^2)^{1/2}$ as a function of the number of
atoms injected into the cavity in their excited state
and detected afterwards in the same excited state.
Even though for $\Delta\tau\ll\Delta\tau_c$ a convergence
towards a Fock state is still possible, for $\Delta\tau\approx\Delta\tau_c$
such a convergence is completely destroyed: the system escapes every fixed
point because the trapping condition is different for each atom.
\begin{figure}
\centerline{\hbox{\psfig{figure=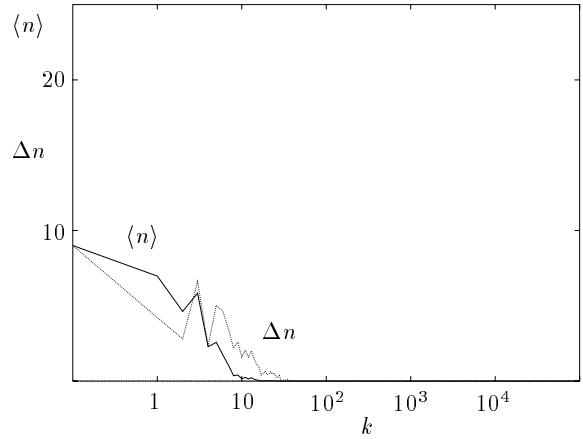,width=3.0in}}}
\vspace{0.2cm}
\caption{Fig.~1. Evolution of $\langle n \rangle$ and $\Delta n$ versus $k$
         (number of atoms) in the scheme of elastic CMs in the case of a 
         large spread ($\Delta\tau=\Delta\tau_c$) in the interaction times.
         The initial coherent state is $|\alpha\rangle = |3\rangle$.}
\label{fg:ee}
\end{figure}
In spite of this, we will show that it is possible to restore the
convergence to fixed points even for large fluctuations
($\Delta\tau > \Delta\tau_c$), if we allow the presence of quantum
interference between the sub-ensembles $|e\rangle\rightarrow |e\rangle$
and $|e\rangle \rightarrow |g\rangle$.

We explain this effect with a simple argument. Let us suppose that
the initial state of each atom is the excited one so that the general
transformation~(\ref{eq:transf}) simplifies to
\begin{equation}
d_n^{(k)} = P_k^{-1/2} \left[ \alpha_k^{(f)*} C_n^{(k)} d_n^{(k-1)}
-i\beta_k^{(f)*}S_{n-1}^{(k)} d_{n-1}^{(k-1)}\right]\;,
\label{eq:transfe}
\end{equation}
where the coefficients of the final atomic
superposition are expressed [9]
in terms of the Rabi frequency $\Omega_k^{(f)}$ and of the interaction
time $T_k^{(f)}$ with the resonant classical field $E_5$ according to
$
\alpha_k^{(f)} = \cos\left(\Omega_k^{(f)}T_k^{(f)}/2\right)
$
and
$
\beta_k^{(f)} = \sin\left(\Omega_k^{(f)}T_k^{(f)}/2\right)e^{i\varphi_f}\;.
$
If we now choose $\varphi_f\simeq -\pi/2$, and neglect the difference
between $n$ and $n-1$, that is $d_n^{(k)}\simeq d_{n-1}^{(k)}$ and
$S_{n-1}^{(k)}\simeq S_n^{(k)}$, Eq.~(\ref{eq:transfe})
approximately reads
\begin{equation}
d_n^{(k)}\simeq P_k^{-1/2}\cos\left(\Omega_k^{(f)}T_k^{(f)}/2-g\tau_k\sqrt{n+1}
\right) d_n^{(k-1)}\; .
\label{eq:transfes}
\end{equation}
Since the atoms (with thermal velocity) cross first the cavity and then
the classical field, $T_k^{(f)}$ and $\tau_k$ in Eq.~(\ref{eq:transfes})
are correlated, even if they are random. This yields a strong suppression
of the fluctuations in the argument of cosine in~(\ref{eq:transfes}).
This is shown in Fig.~2 where we plot (for the scheme
$|e\rangle \rightarrow |e\rangle + |g\rangle$) $\langle n\rangle$ and
$\Delta n$ for fixed interaction times, and for small and large spreads
in the interaction times. Notwithstanding the large fluctuations, the
convergence towards the desired Fock state is still very good. This is
confirmed by the final photon-number distribution $P(n)$, which corresponds to
that of a number state: all the $P(n)$ vanish except one, $P(21)=1$.
\begin{figure}
\centerline{\hbox{\psfig{figure=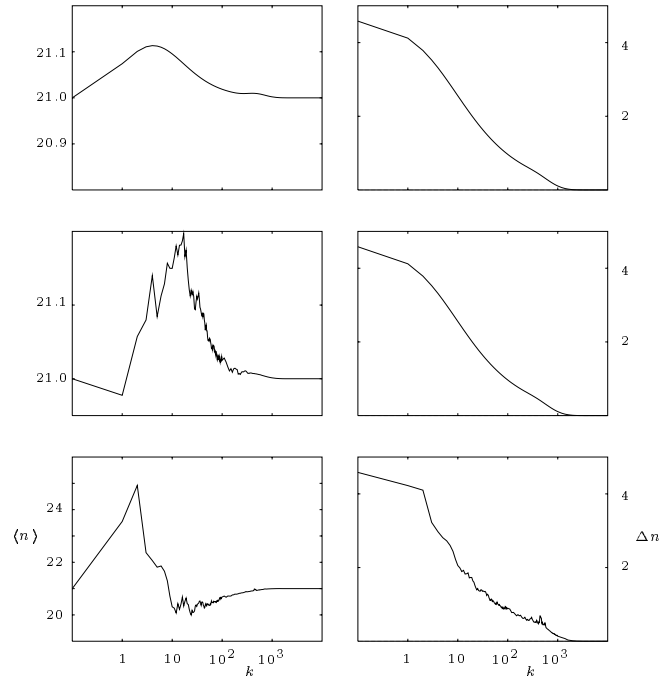,width=3.4in}}}
\vspace{0.2cm}
\caption{Fig.~2. Convergence towards a Fock state---even for a large spread in
         the interaction times---in the scheme $|e\rangle \rightarrow
         |e\rangle + |g\rangle$. Top: behaviour of $\langle n\rangle$ (left)
         and
         $\Delta n$ (right) versus $k$ (number of atoms) in the case of
         fixed interaction times; middle: the same for small fluctuations
         in the interaction times ($\Delta\tau = \Delta\tau_c/10$);
         bottom: the same for large spread in the interaction times
         ($\Delta\tau = 2\Delta\tau_c$). In all cases the initial coherent
         state is $|\alpha\rangle = |\protect\sqrt{21}\rangle$.}
\label{fg:int}
\end{figure}

\kapitola{4. Discussion and Conclusions}

In this paper we have presented a novel scheme which is able to
produce preselected Fock states inside a cavity in which a coherent
state was initially prepared. The final number state is achieved by
sending many two-level atoms in their excited state through the
cavity and by performing a conditional measurement of their internal
degree of freedom in a superposition of the ground and excited states
after they leave the cavity. The proposed scheme---differently from
previous ones [5]--- is quite effective and immune even
to large fluctuations in the interaction times between the atoms and
the cavity field. This is achieved essentially thanks to two basic
ingredients: (a) the conditional measurement of the final state of the
atom, and (b) the quantum interference between the two possible atomic
states ($|e\rangle$ and $|g\rangle$) after the interaction with the
cavity. Since the classical NSMs counterpart of our model is chaotic
(in the regime of random interaction times), and has a quantum dynamics
similar to the classical one, such a striking behaviour
suggests an analogy with recently proposed methods [6] of
controlling classical chaos. These methods, mainly based on classical
feedback, use the extreme sensitivity of chaotic systems to small
perturbations in order to stabilise regular periodic orbits in the
chaotic dynamics. In this perspective, our method can be considered
as a novel (fully quantum) way of stabilising---even for large
fluctuations---fixed points in the quantum dynamics of a system which
is classically chaotic.

\medskip

\noindent {\bf Acknowledgements}

\noindent
We acknowledge helpful and stimulating discussions with I. Averbukh,
R. Chiao, and K. Vogel. M.~F. would like to thank the European
Community (Human Capital and Mobility programme) for support, and
Prof. Gershon Kurizki and his group for the kind hospitality at the
Weizmann Institute of Science, Rehovot, Israel.

\small
\kapitola{References}
\begin{description}
\itemsep0pt
\item{[1]} J.~A.~Wheeler and W.~H.~Zurek (eds.), {\it Quantum Theory and
           Measurement} (Princeton University Press, 1983);
\item{[2]} For previously proposed ways of producing Fock states, see:
           \refer{J.~Krause, M.~O.~Scully, T.~Walther, and H.~Walther}
           {Phys. Rev. A}{39}{1989}{1915}
           J.~M.~ Raimond {\it et al.}, in {\it Laser Spectroscopy IX},
           edited by M.~S.~Feld, J.~E.~Thomas, and A.~Mooradian
           (Academic Press, New York, 1989);
           \refer{J.~R.~Kuklinski}{Phys. Rev. Lett.}{64}{1990}{2507}
\item{[3]} \refer{W.~E.~Lamb}{Physics Today}{22}{1969}{23} for a notable
           example of this approach, see:
           \refer{A.~S.~Parkins, P.~Marte, P.~Zoller, and H.~J.~Kimble}
           {Phys. Rev. Lett.}{71}{1993}{3095}
\item{[4]} The CM approach has been suggested
           by \refer{B.~Sherman and G.~Kurizki}{Phys. Rev. A}{45}{1992}{R7674}
           and developed by \refer{B.~M.~Garraway, B.~Sherman, H.~Moya-Cessa,
           P.~L.~Knight, and G.~Kurizki}{Phys. Rev. A}{49}{1994}{535}
           see also \refer{K.~Vogel, V.~M.~Akulin,
           and W.~P.~Schleich}{Phys. Rev. Lett.}{71}{1993}{1816}
\item{[5]}
\refer{P.~Filipowicz, J.~Javainen, and P. Meystre}
       {J. Opt. Soc. Am. B}{3}{1986}{906}
\item{[6]} The use of small perturbations to control
           chaos has been introduced by \refer{E.~Ott, C.~Grebogi, and
           J.~A.~Yorke}{Phys. Rev. Lett.}{64}{1990}{1196}
           it has been demonstrated experimentally by
           \refer{E.~R.~Hunt}{Phys. Rev. Lett.}{67}{1991}{1953}
           successive developments can be found, {\it e.g.}, in
           \refer{S.~Boccaletti and F.~T.~Arecchi}
           {Europhys. Lett.}{31}{1995}{127}
           for a review, see \refer{T.~Shinbrot,
           C.~Grebogi, E.~Ott, and J.~A.~Yorke}{Nature}{363}{1993}{411}
\item{[7]}
\refer{G.~Harel, G.~Kurizki, J.~K.~McIver, and E.~Coutsias}
      {Phys. Rev. A}{53}{1996}{}
\item{[8]} P.~Meystre, in {\it Progress in Optics XXX}
            E. Wolf (ed.), Elsevier (1992);
\item{[9]} L.~Allen and J.~H.~Eberly, {\it Optical Resonance and Two-Level
            Atoms} (Dover, New York, 1987);
\end{description}
\end{document}